\documentstyle[emulateapj,psfig]{article}
\begin{document}

\title{Spectrum and Duration of Delayed MeV-GeV Emission
of Gamma-Ray Bursts in Cosmic Background Radiation Fields}
\author{Z. G. Dai and T. Lu}
\affil{Department of Astronomy, Nanjing University, Nanjing 210093, China\\
E-mail: daizigao@public1.ptt.js.cn; tlu@nju.edu.cn \\
{\em Accepted for Publication in the Astrophysical Journal on July 30, 2002}}

\begin{abstract}
We generally analyze prompt high-energy emission above a few
hundreds of GeV due to synchrotron self-Compton scattering in
internal shocks. However, such photons cannot be detected because
they may collide with cosmic infrared background photons, leading
to electron/positron pair production. Inverse-Compton scattering
of the resulting electron/positron pairs off cosmic microwave
background photons will produce delayed MeV-GeV emission, which
may be much stronger than a typical high-energy afterglow in the
external shock model. We expand on the Cheng \& Cheng model by
deriving the emission spectrum and duration in the standard
fireball shock model. A typical duration of the emission is $\sim
10^3$ seconds, and the time-integrated scattered photon spectrum
is $\nu^{-(p+6)/4}$, where $p$ is the index of the electron energy
distribution behind internal shocks. This is slightly harder than
the synchrotron photon spectrum, $\nu^{-(p+2)/2}$. The lower
energy property of the scattered photon spectrum is dependent on
the spectral energy distribution of the cosmic infrared background
radiation. Therefore, future observations on such delayed MeV-GeV
emission and the higher-energy spectral cutoff by the {\em
Gamma-Ray Large Area Space Telescope} ({\em GLAST}) would provide
a probe of the cosmic infrared background radiation.
\end{abstract}


%%%%%%%%%%%%%%%%%%%%%%%%%%%%%%%%%
\section {Introduction}
Gamma-ray bursts (GRBs) are the brightest electromagnetic
phenomena in the universe. Their radiation features have been well
understood even though their origin has been unknown. In the
standard fireball shock model (Piran 1999; van Paradijs, Kouveliotou
\& Wijers 2000; M\'esz\'aros 2002), GRBs are explained to be due to
the dissipation of kinetic energy of an expanding fireball with
an average Lorentz factor more than $10^2$ in the internal shocks
produced by collisions between different shells in the fireball.
Afterglows are also considered to be due to the dissipation of kinetic
energy in the external shocks generated by collision of the fireball with
its surrounding medium. Such a model has been supported by a variety of
observations on GRBs and afterglows (e.g., Wijers, Rees \& M\'esz\'aros
1997; Waxman 1997; Vietri 1997). Theoretically, it has also
predicted high-energy emission from GRBs and afterglows
(for a review see M\'esz\'aros 2002). The observed spectra of GRBs
sometimes extend to GeV-TeV photon energies. For example,
the EGRET experiment detected an 18 GeV photon from GRB 940217
(Hurley et al. 1994), and the Milagrito air shower experiment tentatively
detected the TeV emission from GRB 970417a (Atkins et al. 2000). These
GRBs might have originated at a redshift of $z<2$ (Salamon \& Stecker
1998; Totani 2000).

The purpose of this paper is to present a new prediction of
delayed MeV-GeV emission of GRBs in the cosmic background
radiation fields. Such emission arises from inverse-Compton
scattering of electron/positron pairs off cosmic microwave
background (CMB) photons. In \S 2 we generally discuss prompt
high-energy emission above a few hundreds of GeV due to
synchrotron self-Compton scattering in internal shocks based on
the standard fireball model. However, such high-energy photons
cannot be detected because they may be absorbed in interacting
with the cosmic infrared background radiation (Stecker, De Jager
\& Salamon 1992; Madau \& Phinney 1996; MacMinn \& Primack 1996;
Malkan \& Stecker 1998; Salamon \& Stecker 1998). In \S 3 we show
that inverse-Compton scattering of the resulting electron/positron
pairs off CMB photons will produce delayed MeV-GeV emission by
deriving the time-integrated spectrum of such emission and by
estimating the emission duration. In \S 4 we summarize our
findings and discuss their implications. Cheng \& Cheng (1996)
have discussed similar delayed MeV-GeV emission, and here we
further calculate the emission spectrum and estimate the emission
duration in the standard fireball shock model.

%%%%%%%%%%%%%%%%%%%%%%%%%%%%%%%%%
\section{High-Energy Radiation from Internal Shocks}

Let us consider a compact source that produces a wind (i.e., a variable
fireball) with an average luminosity of $L$ and with an average
mass loss rate of $\dot{M}=L/\eta c^2$. Initially, the bulk
Lorentz factor of the wind, $\Gamma$, increases linearly with
radius, until most of the wind energy is converted to kinetic
energy and $\Gamma$ saturates at $\Gamma\sim \eta$. After this
time, fluctuations of $\Gamma$ due to variabilities in $L$ and/or
$\dot{M}$ on time scale $\Delta t$ could lead to internal shocks
in the wind at radius $R\sim \Gamma^2c\Delta t$. We also assume
that the internal shock Lorentz factors $\Gamma_i\sim $ a few in
the wind rest frame, so that the internal shocks could reconvert a
significant fraction of the kinetic energy to internal energy,
which is then emitted as $\gamma$-rays by synchrotron and
inverse-Compton (IC) radiation of the electrons accelerated by the
shocks.

The comoving electron number density of the unshocked wind matter
at radius $R$ is $n'_e=L/(4\pi \Gamma^6 \Delta t^2m_pc^5)$, where
$m_p$ is the proton mass. The internal energy density of the
shocked wind matter is thus approximated by $e'\approx
4\Gamma_i(\Gamma_i-1)n'_e m_pc^2$ (Blandford \& McKee 1976). The
electron energy distribution behind the shocks is usually a power
law: $dn_e/d\gamma_e\propto \gamma_e^{-p}$ for
$\gamma_e\ge\gamma_m$. Assuming that $\epsilon_e$ and $\epsilon_B$
are constant fractions of the internal energy density going into
the electrons and the magnetic field respectively, we obtain the
minimum electron Lorentz factor
$\gamma_m=[(p-2)/(p-1)](m_p/m_e)\epsilon_e\Gamma_i$ (where $m_e$
is the electron mass) and the magnetic field strength of the
shocked wind matter $B=(8\pi \epsilon_B e')^{1/2}=6.2\times 10^3
\epsilon_{B,-2}^{1/2}(\Gamma_i/2)^{1/2}\Gamma_{500}^{-3}L_{52}^{1/2}
\Delta t_{-2}^{-1}$\,G, where
$\epsilon_{B,-2}=\epsilon_B/10^{-2}$, $\Gamma_{500}=\Gamma/500$,
$L_{52}=L/10^{52}{\rm erg}\,{\rm s}^{-1}$, and $\Delta
t_{-2}=\Delta t/10^{-2}{\rm s}$. Thus, we estimate the
characteristic frequency of synchrotron radiation from the
shock-accelerated electrons,
\begin{eqnarray}
\nu_m & = & \Gamma\gamma_m^2 \frac{eB}{2\pi m_ec}\nonumber \\ & =
& 3.0\times 10^{19}[(p-2)/(p-1)]^2\nonumber \\ & & \times \,\,
\epsilon_{e,0.5}^2
\epsilon_{B,-2}^{1/2}(\Gamma_i/2)^{5/2}\Gamma_{500}^{-2}L_{52}^{1/2}
\Delta t_{-2}^{-1}\,{\rm Hz},
\end{eqnarray}
where $\epsilon_{e,0.5}=\epsilon_e/0.5$. According to Sari, Piran
\& Narayan (1998), the cooling Lorentz factor
$\gamma_c\ll\gamma_m$ for typical parameters of the internal shock
model, implying that the shock-accelerated electrons are in the
fast cooling regime. Therefore, the synchrotron spectral
luminosity is
\begin{equation}
L_\nu\equiv \frac{h\nu N_\nu}{\Delta t} = \left \{
       \begin{array}{ll}
         L_{\nu_m}(\nu/\nu_m)^{-1/2} & {\rm if}\,\,\nu<\nu_m\\
         L_{\nu_m}(\nu/\nu_m)^{-p/2} & {\rm if}\,\,\nu\ge\nu_m,
       \end{array}
       \right.
\end{equation}
where $h$ is Planck's constant and $N_\nu$ is the synchrotron photon
number per unit frequency in $\Delta t$.

Although GRBs are optically thin to electron scattering, some
synchrotron photons will Compton scatter on the shock-accelerated
electrons, producing an additional IC component at higher energies
(Papathanassiou \& M\'esz\'aros 1996; Pilla \& Loeb 1998;
Panaitescu \& M\'esz\'aros 2000). The ratio of the IC to
synchrotron radiation luminosity can be estimated by
\begin{equation}
Y\equiv\frac{L_{\rm IC}}{L_{\rm syn}}=\frac{\nu_m^{\rm IC}L_{\nu_m}
^{\rm IC}}{\nu_mL_{\nu_m}}=\left(\frac{\epsilon_e}{\epsilon_B}\right)^{1/2},
\end{equation}
where we have assumed that $\epsilon_e\gg\epsilon_B$
and the radiating electrons are in the fast cooling regime
(Panaitescu \& Kumar 2000; Sari \& Esin 2001;
Zhang \& M\'esz\'aros 2001). Since the characteristic IC frequency
\begin{eqnarray}
\nu_m^{\rm IC} & \simeq & \gamma_m^2\nu_m\nonumber \\ & = &
1.0\times 10^{26}[(p-2)/(p-1)]^4\nonumber \\ & & \times \,\,
\epsilon_{e,0.5}^4\epsilon_{B,-2}^{1/2}(\Gamma_i/2)^{9/2}
\Gamma_{500}^{-2}L_{52}^{1/2}\Delta t_{-2}^{-1}\,{\rm Hz},
\end{eqnarray}
the ratio of the IC spectral luminosity at $\nu_m^{\rm IC}$ to
synchrotron spectral luminosity at $\nu_m$ is
\begin{equation}
\frac{L_{\nu_m}^{\rm IC}}{L_{\nu_m}}=\gamma_m^{-2}Y,
\end{equation}
which is derived from equation (3), and the ratio of the IC to
synchrotron photon number per unit frequency in $\Delta t$ is
further given by
\begin{equation}
\frac{N_\nu^{\rm IC}}{N_\nu}=\gamma_m^{p-2}Y,
\,\,\,\,{\rm for}\,\,\nu\ge\nu_m^{\rm IC}.
\end{equation}

The {\em intrinsic} cutoff energy in the IC component is defined
by ${\cal E}_{\rm cut}^{\rm in}=\min({\cal E}_{\rm KN}^{\rm IC},
{\cal E}_0^{\rm in})$, where ${\cal E}_{\rm KN}^{\rm IC}$ is the
Klein-Nishina limit and  ${\cal E}_0^{\rm in}$ is the energy at
which a photon may be attenuated due to pair production through
interactions with softer photons (also radiated from the internal
shocks) whose frequency is equal to or larger than $\nu_{\rm
an}=(\Gamma m_ec^2)^2 /(h{\cal E}_0^{\rm in})$.
We first estimate ${\cal E}_{\rm KN}^{\rm
IC}$. There is the maximum Lorentz factor $\gamma_M$ in the
electron energy distribution behind the shocks due to the fact
that the shock-acceleration time cannot exceed the cooling time.
This gives  $\gamma_M=\{3 e/[\sigma_T B(1+Y)]\}^{1/2}$  where
$e$ is the electron charge and $\sigma_T$
is the Thomson cross section. Thus, we obtain the
Klein-Nishina limit ${\cal E}_{\rm KN}^{\rm IC}=\gamma_M\Gamma
m_ec^2\simeq 56\epsilon_{e,0.5}^{-1/4}(\Gamma_i/2)^{-1/4}
\Gamma_{500}^{5/2}L_{52}^{-1/4}\Delta t_{-2}^{1/2}\,$TeV. To
compute ${\cal E}_0^{\rm in}$, we adopt an analytical approach
proposed by Lithwick \& Sari (2001). Integrating the
synchrotron spectral luminosity over frequency should lead to the
total synchrotron luminosity, which further gives the synchrotron
photon number per unit frequency at $\nu_m$: $N_{\nu_m}=
\{(p-2)/[2(p-1)]\}\epsilon_e (1+Y)^{-1}(h\nu_m^2)^{-1}L\Delta t$.
The total synchrotron photon number at frequencies larger than
$\nu$ can be estimated by $N_{>\nu}\sim \int_\nu^\infty N_\nu
d\nu= \int_\nu^\infty N_{\nu_m} (\nu/\nu_m)^{-(p+2)/2}d\nu
=\{(p-2)/[p(p-1)]\}\epsilon_e(1+Y)^{-1} (h\nu_m)^{-1}L\Delta
t(\nu/\nu_m)^{-p/2}$. The attenuation optical depth is
\begin{equation}
\tau_{\gamma\gamma}^{\rm in}=\frac{(11/180)\sigma_TN_{>\nu_{\rm an}}}
{4\pi (\Gamma^2c\Delta t)^2}.
\end{equation}
The requirement that $\tau_{\gamma\gamma}^{\rm in}=1$ leads to
the upper energy limit
\begin{eqnarray}
{\cal E}_0^{\rm in} & = & 160\epsilon_{e,0.5}^{-(2p-3)/p}
\epsilon_{B,-2}^{-1/2}(\Gamma_i/2)^{-5(p-2)/(2p)}\nonumber \\ & &
\times \,\,\Gamma_{500}^{4(p+1)/p} L_{52}^{-(p+2)/(2p)} \Delta
t_{-2}\,\,{\rm GeV},
\end{eqnarray}
where the numerical value is quoted for $p=2.2$.  It is easy to
see that ${\cal E}_{\rm KN}^{\rm IC}\gg {\cal E}_0^{\rm in}$ for
typical parameters and thus the intrinsic cutoff energy is in fact
given by equation (8).

\section{Pair Production and Reradiation in Cosmic Background Photon Fields}

High-energy $\gamma$-rays emitted from the internal shocks may not
only be intrinsically attenuated due to pair production through
interactions with softer photons from the same radiation regions
but may also be absorbed in the {\em external} background
radiation fields when these $\gamma$-rays travel towards the
observer. In the latter case, the observed cutoff energy, ${\cal
E}_{\rm cut}^{\rm ob}$, is determined by the redshift ($z$) and
comoving energy, $\varepsilon_\gamma(z)$, of the background
photons that dominate the pair production optical depth
($\tau_{\gamma\gamma}^{\rm ex}$) because the cross section peaks
when ${\cal E}_{\rm cut}^{\rm ob}\varepsilon_\gamma(z) \sim
(m_ec^2)^2/(1+z)$. It is general that at low redshifts the
spectral energy distribution of the starlight background radiation
peaks at infrared wavelengths and thus ${\cal E}_{\rm cut}^{\rm
ob}$ is in the TeV energy range (Salamon \& Stecker 1998).
Detailed predictions of ${\cal E}_{\rm cut}^{\rm ob}$ differ from
model to model, depending on the theoretical treatment adopted for
the stellar emissivity. For example, the observed cutoff energy
when $\tau_{\gamma\gamma}^{\rm ex}=1$ for $z=1$ is $50\,{\rm
GeV}\lesssim {\cal E}_{\rm cut}^{\rm ob}\lesssim 80\,{\rm GeV}$
for different models used in Salamon \& Stecker (1998).

The pair production optical depth, $\tau_{\gamma\gamma}^{\rm ex}$,
strongly depends on the $\gamma$-ray energy (${\cal E}_\gamma^{\rm
ob}$). Salamon \& Stecker (1998) numerically calculated
$\tau_{\gamma\gamma}^{\rm ex}$ as a function of the photon energy
for several fixed redshifts by considering the stellar emissivity with
and without metallicity correction, shown in their Figures 6 and
7, respectively. It can be seen from their Figure 6 that if $z=1$,
then $\tau_{\gamma\gamma}^{\rm ex}\simeq 1$ for ${\cal
E}_\gamma^{\rm ob} =50\,$GeV, but $\tau_{\gamma\gamma}^{\rm
ex}\simeq 10$ for ${\cal E}_\gamma^{\rm ob}= 300\,$GeV. In this
model of Salamon \& Stecker (1998), therefore, a photon with
energy of ${\cal E}_\gamma^{\rm ob}\gtrsim 300\,$GeV must have a
pair production optical depth $\tau_{\gamma\gamma}^{\rm ex}\gg 1$.
This implies that such high-energy $\gamma$-rays may be locally
attenuated once they are radiated from internal shocks. For
simplicity, we will neglect any redshift correction for
low-redshift sources discussed in this paper.

The resulting electron/positron pairs have Lorentz factors of
$\gamma_e\equiv {\cal E}_\gamma^{\rm ob}/(2m_ec^2)\ge \gamma_{\rm
min}\equiv {\cal E}_0^{\rm ex}/(2m_ec^2)= 3\times 10^5{\cal
E}_{0,300}^{\rm ex}$, where ${\cal E}_0^{\rm ex}={\cal
E}_{0,300}^{\rm ex}\times 300\,{\rm GeV}$ is the minimum energy of
photons that are locally attenuated in the external infrared
background radiation field. The pairs will Compton scatter the CMB
photons. As a result, the initial energy of a microwave photon,
$h\nu_0$, is boosted by IC scattering up to an average value $\sim
\gamma_e^2h\nu_0\ge 57({\cal E}_{0,300}^{\rm ex})^2\,$MeV, where
$h\nu_0=2.7kT$ is the mean energy of the CMB photons with $T\simeq
2.73\,$K and $k$ is the Boltzmann constant. The IC lifetime (in
the local rest frame) of an electron with Lorentz factor of
$\gamma_e$ reads
\begin{equation}
\tau(\gamma_e) = \frac{3m_ec}{4\gamma_e\sigma_Tu_{\rm cmb}}\le
2.4\times 10^{14}({\cal E}_{0,300}^{\rm ex})^{-1}\,\,{\rm s},
\end{equation}
where $u_{\rm cmb}=aT^4$ is the CMB energy density and $a$ is the
radiation constant. It is not difficult to find that the typical
length in which most of the electron energy is lost, $\sim
c\tau(\gamma_e)$, is much less than the distance from the source
to the observer, implying that energy loss of the electron due to
IC scattering is also local. Beyond this length, energy loss of
the electron and its emission become insignificant in the absence
of any acceleration.

The typical duration of the scattered photons after the GRB
trigger in the observer's frame is estimated by $\tau^{\rm
ob}=\max(\tau^{\rm ob}_1,\tau^{\rm ob}_2)$, where $\tau^{\rm
ob}_1$ is the observed IC cooling lifetime,
\begin{equation}
\tau^{\rm ob}_1\simeq\frac{\tau(\gamma_e)}{2\gamma_e^2} \le
1.3\times 10^3({\cal E}_{0,300}^{\rm ex})^{-3}\,\,{\rm s},
\end{equation}
and $\tau^{\rm ob}_2\simeq R_{\rm pair}/(2\gamma_e^2c)$ is
the angular timescale (Piran 1999).
Here $R_{\rm pair}=(0.26\sigma_Tn_{\rm IR})^{-1}\simeq 5.8\times
10^{24}(n_{\rm IR}/1\,{\rm cm}^{-3})^{-1}$ cm is the typical
pair-production radius (where $n_{\rm IR}\sim 1\,{\rm cm}^{-3}$ is
the cosmic infrared photon number density, Protheroe \& Stanev 1993).
Thus, we have
\begin{equation}
\tau^{\rm ob}_2\le 1.0\times 10^3(n_{\rm IR}/1\,{\rm
cm}^{-3})^{-1}({\cal E}_{0,300}^{\rm ex})^{-2}\,\,{\rm s}.
\end{equation}
Therefore, most of the scattered photons will reach the observer
in $\sim 10^3$ seconds. This time is much longer than a
typical GRB duration.

We next derive the scattered photon spectrum. From equations (2),
(3) and (6), we first write the spectrum of the electrons and
positrons as follows
\begin{eqnarray}
\frac{dN_e}{d\gamma_e} & = & 2\gamma_m^{p-2}Y\nu_m N_{\nu_m}
\left(\frac{h\nu_m}{2m_ec^2}\right)^{p/2}
\gamma_e^{-(p+2)/2},\nonumber \\ & & {\rm if}\,\,\,\gamma_{\rm
min}\le\gamma_e\le\gamma_{\rm max},
\end{eqnarray}
where $\gamma_{\rm max}={\cal E}_0^{\rm in}/(2m_ec^2)$. The
scattered photon number per unit frequency is given by (Blumenthal
\& Gould 1970, hereafter BG70)
\begin{equation}
N_\nu^{\rm sc}=\int\int \left(\frac{dN_e}{d\gamma_e}\right)
\left(\frac{dN_{\gamma_e,\epsilon}}{dt'd\nu}\right)\tau(\gamma_e)
d\epsilon d\gamma_e,
\end{equation}
where $dN_{\gamma_e,\epsilon}/dt'd\nu$, expressed by equation
(2.42) of BG70 in the Thomson limit, is the spectrum of photons
scattered by an electron with Lorentz factor of $\gamma_e$ from a
segment of the CMB photon gas of differential number density
$n(\epsilon)d\epsilon$, and $t'$ is the time measured in the local
rest frame. Please note that the scattered photon number per unit
frequency given by equation (13) is time-integrated by multiplying
$\tau(\gamma_e)$ in equation (2.61) of BG70. After integrating
over $\gamma_e$ and $\epsilon$, we obtain the ratio of the
scattered to synchrotron photon number per unit frequency
\begin{eqnarray}
\frac{N_\nu^{\rm sc}}{N_\nu} & = &
\frac{9(m_ec^2)^{-(p-2)/2}}{2^{(p+8)/2}\pi^2 (\hbar
c)^3aT^4}\gamma_m^{p-2}Y\nonumber \\ & & \times \,\,
F(q)(kT)^{(p+14)/4}(h\nu)^{(p-2)/4} \nonumber \\ & = &
2^{-p/2}Y\frac{F(q)}{F(3)}\left[\frac{\gamma_m^4(kT)(h\nu)}
{(m_ec^2)^2}\right]^{(p-2)/4},
\end{eqnarray}
where $F(q)$ is a function of $q\equiv (p+4)/2$,
\begin{equation}
F(q)=\frac{2^{q+3}(q^2+4q+11)}{(q+3)^2(q+1)(q+5)}
\Gamma[(q+5)/2]\zeta[(q+5)/2]
\end{equation}
with the Gamma function and Riemann zeta function.
Therefore, the scattered photon spectrum becomes
\begin{equation}
N_\nu^{\rm sc}\propto \nu^{-(p+6)/4}.
\end{equation}
The lower energy limit of the scattered photons is
\begin{equation}
h\nu_1\sim \gamma_{\rm min}^2(2.7kT)= 57({\cal E}_{0,300}^{\rm
ex})^2\,{\rm MeV},
\end{equation}
while the upper energy limit is
\begin{eqnarray}
h\nu_2 & \sim & \gamma_{\rm max}^2(2.7kT)\nonumber \\ & \simeq &
16\epsilon_{e,0.5}^{-2(2p-3)/p}\epsilon_{B,-2}^{-1}
(\Gamma_i/2)^{-5(p-2)/p}\nonumber \\
& & \times\,\,\Gamma_{500}^{8(p+1)/p} L_{52}^{-(p+2)/p}\Delta
t_{-2}^2\,{\rm MeV},
\end{eqnarray}
which is derived from equation (8).

We now consider a simple example. The typical parameters of the
internal shock model are taken:
$\epsilon_{e,0.5}=\epsilon_{B,-2}=L_{52}= \Delta
t_{-2}=\Gamma_i/2=1$, $\Gamma_{500}=1.4$ and $p=2.2$. Furthermore,
${\cal E}_{0,300}^{\rm ex}=1$ is assumed. Therefore, we obtain
$h\nu_1\sim 57\,$MeV and $h\nu_2\sim 0.8\,$GeV. From equation
(14), we further find $N_\nu^{\rm sc}/N_\nu$ increases from $\sim
5.0$ to $\sim 5.8$ when the observed energy of the scattered
photons increases from $h\nu_1$ to $h\nu_2$. This leads to a
delayed MeV-GeV emission component. It is interesting to note that
the total energy release of such emission is not only comparable
to that of a typical GRB, but also much larger than the total
MeV-GeV energy release in an early afterglow as calculated by
Dermer, Chiang \& Mitman (2000) and Zhang \& M\'esz\'aros (2001).

\section{Discussion and Conclusions}

We have discussed prompt high-energy emission from the internal
shocks, which are produced by collisions between shells with
different Lorentz factors in an ultrarelativistic wind ejected
from the central engine of GRBs. Such emission may arise from
synchrotron self-Compton scattering in the internal shocks. We
derived the intrinsic cutoff energy due to pair production through
interactions of high-energy photons with softer photons from the
same radiation regions. Our intrinsic cutoff energy is consistent
with the observed spectra of low-redshift GRBs such as GRB 940217
and GRB 970417a. However, another cutoff energy appears in the
$\gamma$-ray spectra. This cutoff results from interactions of
high-energy photons with external cosmic infrared background
photons. The electron/positron pairs produced during such
interactions own a significant fraction of the explosion energy
and they will Compton scatter the CMB photons. We expanded on the
Cheng \& Cheng (1996) model by deriving the emission spectrum and
duration in the standard fireball shock model. The resulting
emission will be able to reach the observer in $\sim 10^3$
seconds. The time-integrated scattered photon spectrum is
$N_\nu^{\rm sc}\propto \nu^{-(p+6)/4}$. This is slightly harder
than the internal shock emission spectrum, $N_\nu\propto
\nu^{-(p+2)/2}$. For typical parameters of the internal shock
model, the observed energies of the scattered photons are in the
MeV-GeV range,  and the ratio of the scattered to synchrotron
photon number per unit frequency, $N_\nu^{\rm sc}/N_\nu$, is
typically a few.

Of course, whether or not there is such delayed emission depends
on the parameters of the internal shock model. Some parameters
(e.g., $L$, $\epsilon_e$ and $\epsilon_B$) adopted in this paper
are consistent with detailed fits to the multi-wavelength data of
a few afterglows (Wijers \& Galama 1999; Freedman \& Waxman 2001;
Panaitescu \& Kumar 2002). Two other parameters (e.g., $p$ and
$\Delta t$) are consistent with the BATSE observations (Preece et
al. 2000). The bulk Lorentz factor of the wind, $\Gamma$, is a
crucial parameter because ${\cal E}_0^{\rm in}$ is strongly
dependent of $\Gamma$ (see equation [8]). The observed TeV
emission from GRB 970417a requires that $\Gamma$ be equal to or
larger than 700, {\em provided that the other parameters are given
in the simple example of \S 3}. Furthermore, some arguments on the
optical flash of GRB 990123 in the reverse shock model (Wang, Dai
\& Lu 2000; Soderberg \& Ramirez-Ruiz 2002) show that $\Gamma$ may
be of the order of $10^3$ (but also see Sari \& Piran [1999], who
suggested that $\Gamma$ in this burst could be about 200). If
$\Gamma$ is close to or larger than 600, we can indeed see a
delayed MeV-GeV emission spectrum. If $\Gamma<600$, however, the
intrinsic cutoff energy could be less than the external
pair-production minimum energy at which a photon is locally
attenuated through interaction with the infrared background
radiation, so that there could not be the delayed MeV-GeV emission
discussed here.

It is seen from equations (6) and (14) that the Compton parameter
$Y$ determines the TeV emission from synchrotron self-Compton
scattering in internal shocks, and thus the spectral photon number
of the delayed MeV-GeV emission. In addition, additional TeV
photons produced by other mechanisms, e.g., photo-pion and
inelastic proton-neutron collisions (Waxman \& Bahcall 1997;
B\"ottcher \& Dermer 1998; Derishev, Kocharovsky \& Kocharovsky
1999; Bahcall \& M\'esz\'aros 2000; Dai \& Lu 2001), and
synchrotron self-Compton scattering both in external reverse
shocks (Wang, Dai \& Lu 2001a, 2001b) and in early-time forward
shocks (Dermer, Chiang \& Mitman 2000; Zhang \& M\'esz\'aros
2001), may be absorbed in the cosmic infrared photon field,
leading to electron/positron pairs. The inverse Compton scattering
of such pairs off CMB photons may have a non-negligible
contribution to the delayed MeV-GeV emission.

The MeV-GeV emission studied in this paper will be detectable by
next-generation $\gamma$-ray satellites such as the {\em Gamma-Ray
Large Area Space Telescope} ({\em GLAST}). In the {\em
Swift-GLAST} era, a plenty of GRBs localized by the {\em Swift}
satellite are expected to be detected by {\em GLAST}. Such
observations, particularly on the delayed MeV-GeV emission and
higher-energy spectral cutoff, may provide a probe of the cosmic
infrared background radiation. This in turn may help to constrain
some of the most fundamental uncertainties in physical models of
the star formation.

%%%%%%%%%%%%%%%%%%%%%%%%%%%%%%%%%
\acknowledgments

We are very grateful to the referee for helpful comments and
criticism that enabled us to improve the manuscript significantly,
and to Y. F. Huang, S. Kobayashi, X. Y. Wang, D. M. Wei and B.
Zhang for valuable discussions. This work was supported by the
National Natural Science Foundation of China (grants 19825109 and
19973003) and the National 973 Project (NKBRSF G19990754).

\end{document}